\documentclass{cpcauth}

\usepackage{amssymb}

\begin{document}

\begin{frontmatter}



\title{ Generalized Ensemble Simulations for Complex Systems }


\author{Bernd A. Berg {\small (berg@hep.fsu.edu)} }

\address{ Department of Physics, Florida State University, 
Tallahassee FL 32306, USA}

\begin{abstract}
 The most efficient MC weights for the calculation of physical, canonical 
 expectation values are not necessarily those of the canonical ensemble. 
 The use of suitably generalized ensembles can lead to a much faster
 convergence of the simulation. Although not realized by nature, these 
 ensembles can be implemented on computers.
 In recent years generalized ensembles have in particular
 been studied for the simulation of complex systems. For these
 systems it is typical that conflicting constraints lead to free
 energy barriers, which fragment the configuration space. Examples
 of major interest are spin glasses and proteins. In my overview I
 first comment on the strengths and weaknesses of a few major
 approaches, multicanonical simulations, transition variable methods,
 and parallel tempering. Subsequently, two applications are presented:
 a new analysis of the Parisi overlap distribution for the 3d 
 Edwards-Anderson Ising
 spin glass and the helix-coil transition of amino-acid homo-oligomers.
\end{abstract}

\end{frontmatter}

\section{Introduction} \label{sec_intro}

Markov chain Monte Carlo (MC) simulations are an indispensable tool for 
the study of physical models. To calculate expectation values in the 
Gibbs canonical ensemble, most simulations use Boltzmann weights. 
However, in the course of time it has 
become clear that the Boltzmann factor is not always
the most efficient weight to generate the desired
configurations of the canonical ensemble. In particular before
embarking on a large-scale computer simulation, one of the questions 
which ought to be addressed is ``What are suitable weight factors
for the problem at hand?'' Nowadays simulations of more general
ensembles are employed for an increasing number of
applications in physics, chemistry, structural biology and
other areas, for reviews see~\cite{Be00,HaOk99,MiSu01,WaSw01}. 

For instance, the multicanonical (MUCA)
method~\cite{ToVa77,BeNe92,BeCe92,HaOk93,Be96,WaLa01}
calculates in one simulation the canonical ensemble at many temperatures.
Its best-established application is calculations of interface tensions 
at first order phase transitions. There, canonically rare configurations 
are simply enhanced, the {\it statical} case. This is similar to the 
importance sampling idea, which for the calculation of canonical 
expectation values led from naive sampling to the Markov process 
sampling with the Boltzmann weights. The additional complication is now 
that the new weights are {\it a priori} unknown. The nowadays most 
important applications of the MUCA method are for complex
systems~\cite{BeCe92,HaOk93}. There, the rational of MUCA
simulations is to flatten free energy barriers, such that the 
{\it dynamics} of the simulation becomes improved. The enhancement of 
particular configurations can no longer be explicitly controlled and the 
issue of optimizing MC algorithms for complex systems is far from 
being well understood, although some improvements have been quite
spectacular.

In the next section we introduce the MUCA method, using as
illustration the $2d$ ten-state Potts model with its strong 
first-order phase transition. Subsection~\ref{sec_recursion} 
discusses the weight recursions of Ref.\cite{Be00,Be96} 
and~\cite{WaLa01}.  Random walk and transition matrix
methods~\cite{Be93,SmBr95,Ol96,BeHa98,Ol98,WaTa99,Wa99,MuHe99}
generalize the MUCA approach and are summarized in 
section~\ref{sec_RWTM}. Distinct from these methods are 
multiple Markov chain algorithms~\cite{Ge91,HuNe96,TeJa96} 
(parallel tempering). They are introduced in section~\ref{sec_PT}, 
where their recent combinations~\cite{SuOk00} with MUCA methods
are also sketched.
In section~\ref{sec_complex} we turn to applications for 
complex systems.  Instead of being exhaustive (see the  
reviews~\cite{Be00,HaOk99,MiSu01}) we focus on two recent studies: 
a new analysis of the Parisi overlap distribution for the 3d 
Edwards-Anderson Ising spin glass~\cite{BeBi01} and the helix-coil
transition of the polyalanine protein~\cite{HaOk99a,MiOk00}. 
Outlook and conclusions are given in section~\ref{sec_out}. 

\section{Multicanonical Simulations} \label{sec_muca}

Multicanonical simulations are best established for investigations 
of first-order phase transitions. For them
pseudocritical points $\beta^c (L)$, where $L$
is the lattice size, exist such that the canonical energy 
density $P(E)=P(E;L)$ becomes double-peaked and
the maxima at $E^1_{\max} < E^2_{\max}$ are of equal height
$P_{\max} = P (E^1_{\max}) = P (E^2_{\max})$. In between these
values a minimum is located at some energy $E_{\min}$.
Configurations at $E_{\min}$ are exponentially suppressed like
\begin{equation} \label{Pmin}
P_{\min} = P (E_{\min}) = c_f\, L^p\, \exp ( - f^s A )
\end{equation}
where $f^s$ is the interface tension and $A$ is the minimal area 
between the phases, $A=2L^{d-1}$ for an $L^d$ lattice, $c_f$ and
$p$ are constants. To determine the interface tension with Binder's
histogram method, one has to calculate the quantities
\begin{equation} \label{Itension}
f^s (L) = - {\ln R(L) \over A(L)} ~~{\rm with}~~ 
R(L) = {P_{\min} (L) \over P_{\max} (L) }
\end{equation}
and to make a finite size scaling (FSS) extrapolation of $f^s(L)$ 
for $L\to\infty$. However, for large systems a canonical MC
simulation will practically never visit 
configurations at energy $E=E_{\min}$ and estimates of the ratio 
$R(L)$ will be very inaccurate. The terminology \textit{supercritical
slowing down} was coined to characterize thus an exponential
deterioration of simulation results with lattice size. The MUCA
method overcomes this problem by sampling, in an
appropriate energy range, with an \textit{approximation}
\begin{equation} \label{wmuca}
 \widehat{w}_{mu}(k)=w_{mu}(E^{(k)}) 
       = e^{-b(E^{(k)})\, E^{(k)} + a (E^{(k)})}
\end{equation}
to the inverse density of states  $1/n(E^{(k)})$.
Here the function $b(E)$ is the inverse \textit{microcanonical 
temperature} at energy $E$ and $a(E)$ is the dimensionless,
microcanonical free energy. One samples 
instead of the canonical energy
distribution $P(E)$ a new MUCA distribution
\begin{equation} \label{Pmuca}
 P_{mu} (E) = c_{mu}\, n(E)\, w_{mu} (E) \approx c_{mu}\ .
\end{equation}
The desired canonical distribution is obtained by re-weighting, which 
is rigorous, because the weights $w_{mu}(E)$ used in the actual 
simulation are exactly known. With the approximate relation
(\ref{Pmuca}) the average number of configurations sampled does no
longer depend strongly  on the energy and accurate estimates of the
ratio (\ref{Itension}) $R(L)=P_{\min}/P_{\max}$ become possible.

The MUCA method requires two steps

\begin{enumerate}

\item Obtain a \textit{working estimate} of  the weights (\ref{wmuca}).
      Working estimate means that the approximation to the inverse
      density of states has to be good enough to ensure movement
      in the desired energy range.

\item Perform a Markov chain MC simulation with the weights
      $\widehat{w}_{mu}(k)$. Canonical expectation values are found 
      by re-weighting to the Gibbs ensemble and standard jackknife
      methods allow reliable error estimates.

\end{enumerate}

\begin{figure}[t]
 \begin{picture}(150,155)
    \put(0, 0){\includegraphics{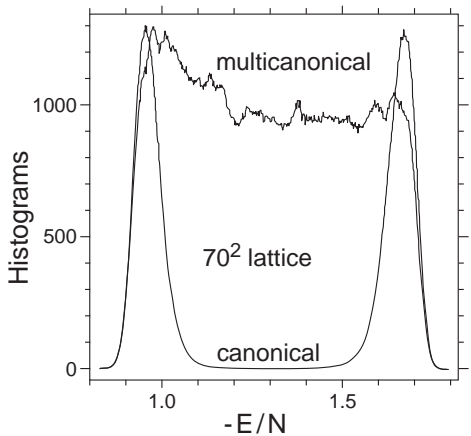}}
  \end{picture}
\caption{ The multicanonical energy histogram $P_{mu}(E)$ together with
its canonically re-weighted energy density $P(E)$ for the $2d$ 10-state
 Potts model on a $70\times 70$ lattice~\cite{BeNe92}.} \label{fig_BN92}
\end{figure}

For the $2d$ 10-state Potts model figure~\ref{fig_BN92} reproduces a
thus obtained MUCA histogram together with its canonically
re-weighted energy histogram~\cite{BeNe92}.
The canonical energy density of this figure gives
the estimate $2 f^s(L)$ of the interface tension on the $70\times 70$
lattice. Combining the such obtained estimates from several lattices
allows the infinite volume FSS extrapolation $2 f^s = 0.0978 (8)$.
At the time of its publication this number was in strong
disagreement with other numerical results. 
Afterwards the exact value was discovered to be 
$2f^s = 0.094701...$~\cite{BoJa92}. This success helped the
acceptance of the method, which by now 
has been applied to a large number of models~\cite{Be00}.

There exist many variants of the MUCA method. 
MUCA refers to calculations of canonical expectation 
values for a temperature range and re-weighting has to be done in the 
internal energy. Similarly, other physical quantities can be considered, 
{\it e.g.} {\it multimagnetical} \cite{BeHa93} refers to simulations
which give results for a certain range of the magnetic field. A variant
for cluster updates due to Janke and Kappler~\cite{JaKa95} is called
{\it multibondic}. In Ref.\cite{BeJa98} the {\it multi-overlap}
algorithm was introduced, which focuses on the Parisi order parameter
of spin glasses (see section~\ref{sec_complex}). Combining MUCA with
multigrid methods has been explored by Janke and Sauer~\cite{JaSa94}.
For molecular dynamics, Langevin and hybrid MC variants see Hansmann
et al.~\cite{HaOkEi96} and, with emphasise on lattice gauge theory,
Arnold et al.~\cite{ArLi98}.

\subsection{Multicanonical Recursion} \label{sec_recursion}

For spin systems with first-order phase transitions the 
FSS behavior is relatively well known. Provided the steps
between system sizes are not too large, it is then possible 
to obtain working estimates of the $w_{mu}(E)$ weights by 
means of a FSS extrapolation from the already simulated smaller 
systems \cite{BeNe92}. However, this method cannot be applied
to complex systems and to get a working estimate of the MUCA
parameters by means of a recursion is then at the heart of the method.

As the overall normalization is irrelevant for updating purposes, it 
is sufficient to consider 
nearest neighbor (in energy) ratios of the weights~(\ref{Pmuca}) 
\begin{equation} \label{w_ratio}
 R^n (E) = {w(E)\over w(E+\epsilon)} = e^{\epsilon b^n (E)}
\end{equation}
where $\epsilon$ is the smallest available energy step and
$n$ refers to the $n^{th}$ iteration.  Initially, $w^0 (E)=1$ is
a good starting value, because the system can then move freely. The
final equation for the recursion, as derived in~\cite{Be00}, reads
\begin{equation} \label{recursion}
 R^{n+1} (E) = R^n (E)\, \left[ {\hat{H}^n (E+\epsilon) \over
 \hat{H}^n (E) } \right]^{\hat{g}^n_0(E)} 
\end{equation}
where the exponent $\hat{g}^n_0(E)$ is recursively determined by the
equations
\begin{equation} \label{g_normalized}
 \hat{g}^n_0 (E) = {g^n_0(E) \over g^n(E) + g^n_0 (E)}\ ,
\end{equation}
\begin{equation}
g^n_0 (E) = {H^n (E+\epsilon)\ H^n (E) \over
H^n (E+\epsilon) + H^n (E)} ~~~~{\rm and}
\end{equation}
\begin{equation}
g^{n+1} (E) = g^n(E) + g^n_0(E),\ g^0(E)=0\, .
\end{equation}
Typically, we want to cover an energy range
$$ E_{\max} - E_{\min} \sim V $$
The optimum for a flat energy distribution is given by a 
random walk in the energy. This implies a CPU time 
increase $ \sim V^2 $
to keep the number of $E_{\max}\to E_{\min}\to E_{\max}$
transitions constant.  The recursion (\ref{recursion})
needs an additional $\sim V^{0.5}$ (optimum) number of
attempts to cover the entire range.
For the two-dimensional Ising model on $L\times L$ lattices, with
$E_{\min}=-2L^2$ (the ground state) and $E_{\max}=0$ (the average
energy at $\beta =0$), table~\ref{tab} gives the number of sweeps 
per transition. The first transition is tedious to find, 
additional transitions are much faster.

\begin{table} \centering
\caption{Sweeps per $E_{\max}=0\to E_{\min}=-L^2\to E_{\max}=0$ 
transition for $L\times L$ Ising models. Here $\tau_1$ refers to
the first transition during the recursion~(\ref{recursion}), 
$\tau_{\mathrm rec}$ to the subsequent transitions during the
recursion and $\tau_{\mathrm prod}$ to the transitions during the
production part.}
\renewcommand{\arraystretch}{1.4}
\setlength\tabcolsep{5pt}
\begin{tabular}{llll}
\hline\noalign{\smallskip}
$L$ & $\tau_1$ & $\tau_{\mathrm rec}$ & $\tau_{\mathrm prod} $\\
\noalign{\smallskip}
\hline
\noalign{\smallskip}
 10 &    1654    (66) &    576   (14) &    545   (14) \\
 20 &   19573   (730) &   3872  (120) &   3855  (110) \\
 30 &   79743  (2600) &  12820  (350) &  11428  (300) \\
 40 &  192274  (6900) &  25901  (820) &  27898 (1700) \\
 50 &  472103 (13000) &  48227 (1600) &  45803 (2300) \\
 60 &  810042 (23000) &  78174 (2300) &  78288 (7200) \\
 80 & 2917722 (56000) & 178456 (8900) & 184236 (9100) \\
\hline
\end{tabular} \label{tab} \end{table}

A new, possibly more efficient, recursion was recently proposed by
F. Wang and Landau~\cite{WaLa01} (my interpretation of the merits
of the method differs somewhat from the authors). They perform
updates with estimators $n(E)$ of the density of states
\begin{equation} \label{WL_updates}
 p(E_1\to E_2) = \min \left[ {n(E_1)\over n(E_2)}, 1\right]\ .
\end{equation}
Each time an energy level is visited, the estimator is
updated according to
\begin{equation} \label{WL_recursion1}
 n(E) \to n(E)\,f 
\end{equation}
where, initially, $n(E)=1$ and $f=f_0=e^1$. Once the
desired energy range is covered, the factor $f$ is refined,
\begin{equation} \label{WL_recursion2}
 f_1=\sqrt{f},\ f_{n+1}=\sqrt{f_{n+1}}\ ,
\end{equation}
until some small value like $f=e^{-8}=1.00000001$ is
reached. A fast convergence towards a good estimator
of the spectral density is reported. Afterwards the usual,
MUCA production runs may be carried out.
A comparison of the recursions~(\ref{recursion}) 
and~(\ref{WL_recursion1},\ref{WL_recursion2}) is subtle, because
in neither case is the optimal schedule or termination point
known.

\section{Transition Matrix Methods} \label{sec_RWTM}

The basic idea of the 
approach~\cite{Be93,SmBr95,Ol96,BeHa98,Ol98,WaTa99,Wa99,MuHe99}
is to employ general transition matrix elements instead of
just weight factors. An example is Wang's~\cite{Wa99} random 
walk algorithm, which uses the transition 
probabilities~\cite{Be93,Ol96,BeHa98,Ol98} 
\begin{equation} \label{rw_transition}
 p(k\to k') = \min \left( 1, {N(E+\triangle E, -\triangle E)\over
                              N(E,\triangle E)} \right)
\end{equation}
where $N(E,\triangle E)$ is the microcanonical average for the
number of transitions from the configuration $k$ with energy $E$
to the configuration $k'$ with energy $E'=E+\triangle E$. A flat
histogram is obtained and the configuration dependence is reduced
to energy dependence. The density of states is obtained from the
equation~\cite{Ol96,BeHa98,Ol98}
\begin{equation} \label{sp_eqn}
 n(E)\, N(E,\triangle E)  = 
\end{equation}
$$ n(E+\triangle E)\, N(E+\triangle E, -\triangle E)\ . $$
The history and general merits of the approach are reviewed 
in~\cite{WaSw01}. There, it is claimed that, in some situations,
equation~(\ref{sp_eqn}) works more efficiently than the Wang-Landau
recursion~(\ref{WL_recursion1},\ref{WL_recursion2}). However,
the issue may not be settled. In particular, it seems~\cite{Wa99} 
that in the random walk MC the use of estimators for 
$N(E,\triangle E)$ faces more serious problems than the use of 
estimators for the weights in MUCA simulations. After defining
reasonable schedules and termination points, one could compare 
the transition times of table~\ref{tab} with those achieved with 
Wang's random walk approach. 

\section{Parallel Tempering} \label{sec_PT}

The developments sketched above have to be distinguished from
the method of multiple Markov chains ({\it parallel tempering}),
which was introduced by Geyer~\cite{Ge91} and, independently, by
Hukusima and Nemoto~\cite{HuNe96}. The latter work was influenced by
the {\it simulated tempering}~\cite{MaPa92} method, which in turn can
be understood as a special case of the {\it method of expanded
ensembles}~\cite{LyMa92}. Parallel tempering is particularly well
suited for parallel processing on clusters of workstations.

Parallel tempering performs $n$ canonical MC simulations
at different $\beta$-values with Boltzmann weight factors
\begin{equation} \label{PT_weights}
w_{B,i} (E^{(k)}) = e^{-\beta_i E^{(k)}},\ i=1, \dots , n
\end{equation}
with $\beta_1 < \beta_2 < ... < \beta_{n-1} < \beta_n$
and allows exchange of neighboring $\beta$-values
\begin{equation} \label{del_beta_PT}
 \beta_{i-1} \longleftrightarrow \beta_i\ ~~{\rm for}~~
 i=2, \dots , n\ .
\end{equation}
These transitions lead to the energy change
\begin{eqnarray} \label{delE_PT}
~~~~ \triangle E\ & = &\ \left( -\beta_{i-1} E^{(k)}_i
  -   \beta_i E^{(k')}_{i-1} \right) \\ \nonumber
& - &\ \left( -\beta_i E^{(k)}_i
  -   \beta_{i-1} E^{(k')}_{i-1} \right) \\ \nonumber
& = &\ \left( \beta_i-\beta_{i-1} \right)\,
      \left( E^{(k)}_i - E^{(k')}_{i-1} \right)
\end{eqnarray}
which is accepted or rejected according to the Metropolis
algorithm. The $\beta_i$-values have to be determined
such that a reasonably large acceptance rate is obtained
for the $\beta$ exchange (\ref{del_beta_PT}) and
Ref.\cite{HuNe96} employs a recursive method.

{\it Remark:} The method works for dynamical but not for statical
supercritical slowing down, because each member of the discrete
set of weight factors samples still a Boltzmann distribution
({\it e.g.} it is not a valid method for calculating
interfacial tensions). This can possibly be overcome by
applying the parallel tempering idea to Gaussian distributions
instead of the canonical ensemble (communicated to the
author by T. Neuhaus).  

Ref.\cite{SuOk00} explores earlier introduced~\cite{Ha97}
combinations of parallel tempering,
called replica-exchange method in~\cite{SuOk00}, with MUCA methods.
The replica-exchange MUCA algorithm performs a short
replica-exchange simulation to determine the MUCA weight
factors and continues with the production part of a regular MUCA
simulation, whereas MUCA replica-exchange works with replicas of 
the MUCA ensemble in different energy ranges.

\section{Applications to Complex Systems} \label{sec_complex}

The relevance of MUCA methods for complex systems was pointed out 
early~\cite{BeCe92}. In these systems one encounters large free 
energy barriers due to disorder and frustration. Multicanonical 
simulations try to overcome the barriers through excursions into 
the disordered phase. In addition, one may try to identify and 
enhance relevant rare configurations.  Here, we limit our interest 
to two recent studies of (a)~spin glasses and (b)~proteins. For 
reviews of applications to complex systems see~\cite{Be00,HaOk99,MiSu01}. 

\subsection{Spin glasses}

\begin{figure}[t]
 \begin{picture}(150,155)
    \put(0, 0){\includegraphics{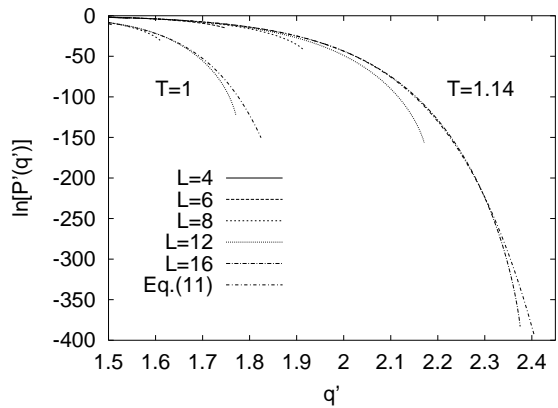}}
  \end{picture}
\caption{Tails of the rescaled overlap~(\ref{Parisi_q}) probability 
density in comparison with predictions of extreme order
statistics [Eq.(11)], see~\cite{BeBi01} ($T=1.14$ is at the freezing
temperature).} \label{fig_spggumble}
\end{figure}

The Parisi overlap parameter $q$ is defined as the overlap of two
replicas of the system at identical temperatures
\begin{equation} \label{Parisi_q}
 q\ =\ \sum_{i=1}^N s_i^1\, s_i^2\ .
\end{equation}
The free-energy structure of the spin glass system is thought to be
intimately related to the barriers in the Parisi order parameter
distributions. The {\it multi-overlap} variant~\cite{BeJa98} of the
MUCA method enhances the minima of the Parisi overlap parameter 
densities. Using the method in large-scale simulations led to a variety 
of new insights, for instance about the self-averaging properties of 
free energy barriers~\cite{BeBi00}. The enhancement is most dramatic
for the tails of the averaged Parisi order parameter distribution.
Recently, this allowed us to demonstrate a conjectured relation between 
the overlap probability density and extreme order statistics over 
about 80 orders of magnitude~\cite{BeBi01}, see figure~\ref{fig_spggumble}.

\subsection{Proteins}

Proteins are linear polymers with the 20 naturally occurring amino 
acids as monomers. Chains smaller than a few tens of amino acids
are called peptides.
The problem is to predict the folded conformation of proteins and
peptides solely from their amino acid sequence. For many years the
emphasis of numerical investigations has been on finding the global
minimum potential energy and the major difficulty encountered is 
the multiple minima problem. Molecular dynamics has been the numerical
method of first choice, but the fraction of stochastic investigations 
shows an increasing trend.

The major advantage of MUCA and related methods in the 
context of proteins is that they allow for investigations of the 
{\it thermodynamics} of the entire free energy landscape of the
protein. This was realized by Hansmann and Okamoto~\cite{HaOk93}
when they introduced MUCA sampling to the problem of protein folding 
and, slightly later, by Hao and Scheraga~\cite{HaSc94}. Since then 
numerous applications have been performed and the simulations 
have been quite successful for peptides.  By now a quite extensive 
literature exists, which is compiled in~\cite{HaOk99,MiSu01}. In
Ref.\cite{YaCe00} the weight recursion~(\ref{recursion}) is adapted
to these systems with continuous energy.
 
\begin{figure}[t]
\vskip -50pt
 \begin{picture}(150,155)
    \put(0, 0){\includegraphics{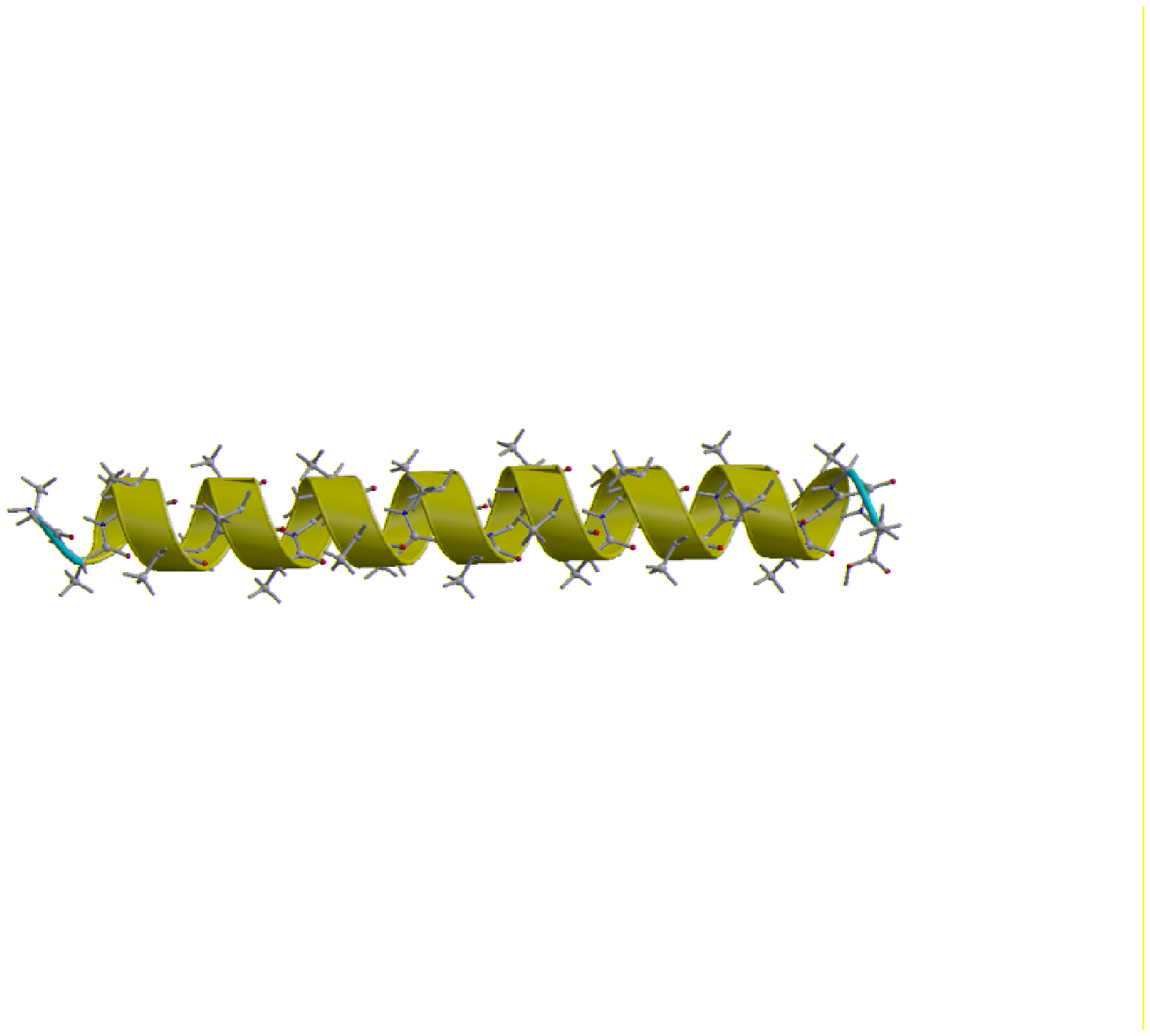}}
  \end{picture}
\caption{ A helix configuration from a multicanonical simulation
of polyalanine~\cite{HaOk99a} (courtesy Ulrich Hansmann and Yuko
Okamoto).} \label{fig_helix}
\end{figure}

A particularly nice application is the helix formation of polyalanine 
by means of a MUCA simulation~\cite{HaOk99a,AlHa00}, depicted in 
figure~\ref{fig_helix}. Up to $N=30$ amino acids are studied 
numerically and a phase transition develops at $T_c=(541\pm 8)\,K$ in 
the infinite volume limit (pseudocritical temperature for finite 
systems are in the range from $427\,K\ (N=10)$ to $518\,K\ (N=30)$). 
FSS estimates of the critical indices $\alpha$, $\nu$ and $\gamma$ 
are consistent with a first-order transition. Progress is made 
concerning the inclusion of aqueous solutions into the 
simulation~\cite{MiOk00}.

\section{Outlook and Conclusions} \label{sec_out}

Sampling of broad energy distributions allows supercritical slowing 
down to be overcome. This is well established for first-order phase
transitions. Systems with conflicting constraints remain, despite some
progress, notoriously difficult and for them most hope lies in
achieving further algorithmic improvements. Reasonably large peptides,
at the border to proteins, are promising systems for a complete
coverage. Transition matrix extensions of the MUCA method and its
combinations with parallel tempering appear promising.

This work was, in part, supported by the U.S. Department of Energy under
contract DE-FG02-97ER41022.

%\bibliography{general}
\end{document}